\documentclass[journal=nalefd,manuscript=letter]{achemso}

\usepackage[version=3]{mhchem}

\author{Jisook Hong}
 \affiliation
 {The Molecular Foundry, Lawrence Berkeley National Laboratory, Berkeley, CA 94720, USA}

\author{David Prendergast}
 \affiliation
 {The Molecular Foundry, Lawrence Berkeley National Laboratory, Berkeley, CA 94720, USA}
 \email{dgprendergast@lbl.gov}

\author{Liang Z. Tan}
 \affiliation
 {The Molecular Foundry, Lawrence Berkeley National Laboratory, Berkeley, CA 94720, USA}
 \email{lztan@lbl.gov}

\title{Layer Edge States Stabilized by Internal Electric Fields in Two-dimensional Hybrid Perovskites}


\begin{document}

\begin{abstract}

  Two-dimensional (2D) organic-inorganic hybrid perovskites have been intensively explored for recent years, due to their tunable band gaps and exciton binding energies, and increased  stability with respect to three-dimensional (3D) hybrid perovskites. There were fascinating experimental observations suggesting the existence of localized edge states in 2D hybrid perovskites which facilitate extremely efficient electron-hole dissociation and long carrier lifetimes. The observations and explanations of the edge states are not quite converging implying that there can be multiple origins for the edge state formation. Using first principles calculations, we demonstrate that layer edge states are stabilized by internal electric fields created by polarized molecular alignment of organic cations in 2D hybrid perovskites when they are two layers or thicker. Our study gives a simple physical explanation of the edge state formation, and it will pave the way for designing and manipulating layer edge states for optoelectronic applications.

\end{abstract}


Organic-inorganic hybrid perovskites are attractive photovoltaic materials because of their high absorption coefficient \cite{Wolf14}, carrier mobility \cite{Oga14}, and long carrier diffusion lengths \cite{Stranks13,Xing13,Shi15,Dong15}. As a result of intensive research, a remarkable improvement in solar cell efficiency from 3.8\% to 29.1\% \cite{Kojima09,NREL} has been achieved during the past decade.  
As it has emerged that inherent instabilities of three-dimensional (3D) hybrid perovskites to moisture, light and heat are detrimental to actual usage in light harvesting applications, two-dimensional (2D) hybrid peroskites have recently been suggested as stable alternatives for use in high-performing long-lasting optoelectronic devices such as solar cells \cite{Smith14,Cao15,Tsai16} and light-emitting diodes (LEDs) \cite{Tsai18,Yang18}.

Dimensional reduction of hybrid perovskites to 2D is achieved by inserting of long insulating organic chains between adjacent inorganic layers.
The general formula of 2D hybrid peroskites is A$_n$A'$_{n-1}$M$_n$X$_{3n+1}$, where A is a large aliphatic or aromatic substituted ammonium cation; A' is a small organic cation such as methylammonium (MA) or formamidium (FA); M is a divalent metal cation; X is a halide anion.
In 2D hybrid perovskites, neighboring inorganic layers are electrically decoupled by insulating organic spacers A, resulting in self-assembled, $n$ octahedral layers thick quantum well structures. 2D hybrid perovskites of thicknesses varying from a monolayer to seven layers have been synthesized in pure phases \cite{Billing07,Stoumpos16,soe19}, and demonstrated promising environmental stability and photostability \cite{Tsai16,Y.Yang18}. 
Their tunability of thickness and chemical composition enables control over optical properties such as band gaps and exciton binding energies. As a result of quantum confinement and reduced dielectric screening effects, the band gap and exciton binding energy of 2D hybrid perovskites increase as the thickness $n$ decreases \cite{Umebayashi03}.  

Experimentally, lower-energy layer edge states were observed in 2D perovskite flakes (BA)$_2$A'$_{n-1}$Pb$_n$Y$_{3n+1}$ (BA = butylammonium; A' = MA, FA; Y = Br, I) when $n\geq2$ or 3 \cite{Blancon17,Shi19,Zhao19}. These long-lived, localized edge states are thought to participate in exciton dissociation pathways, leading to longer carrier lifetimes, and their use in photodetectors \cite{Feng18}. To control and harness these layer edge states, it is crucial to understand their origin and dependence on atomic/molecular structure. While various intrinsic or extrinsic mechanisms for their formation have been proposed, including structural reorganization \cite{Kepenekian18}, loss of BA ligands \cite{Zhang19} and formation of 3D perovskite on edges \cite{Qin20}, it is not clear if a single mechanism is dominant.

In this work, we show that layer edge states are stabilized by the alignment of organic cations A'. Our first-principles density functional theory (DFT) simulations show that this internal electric field leads to charge separation at the edges. This mechanism stabilizes layer edge states for 2D hybrid perovskites with $n\geq2$ layers, but not for $n=$ 1 perovskites, which do not have small orientable A' organic cations.


\begin{figure*}
\includegraphics{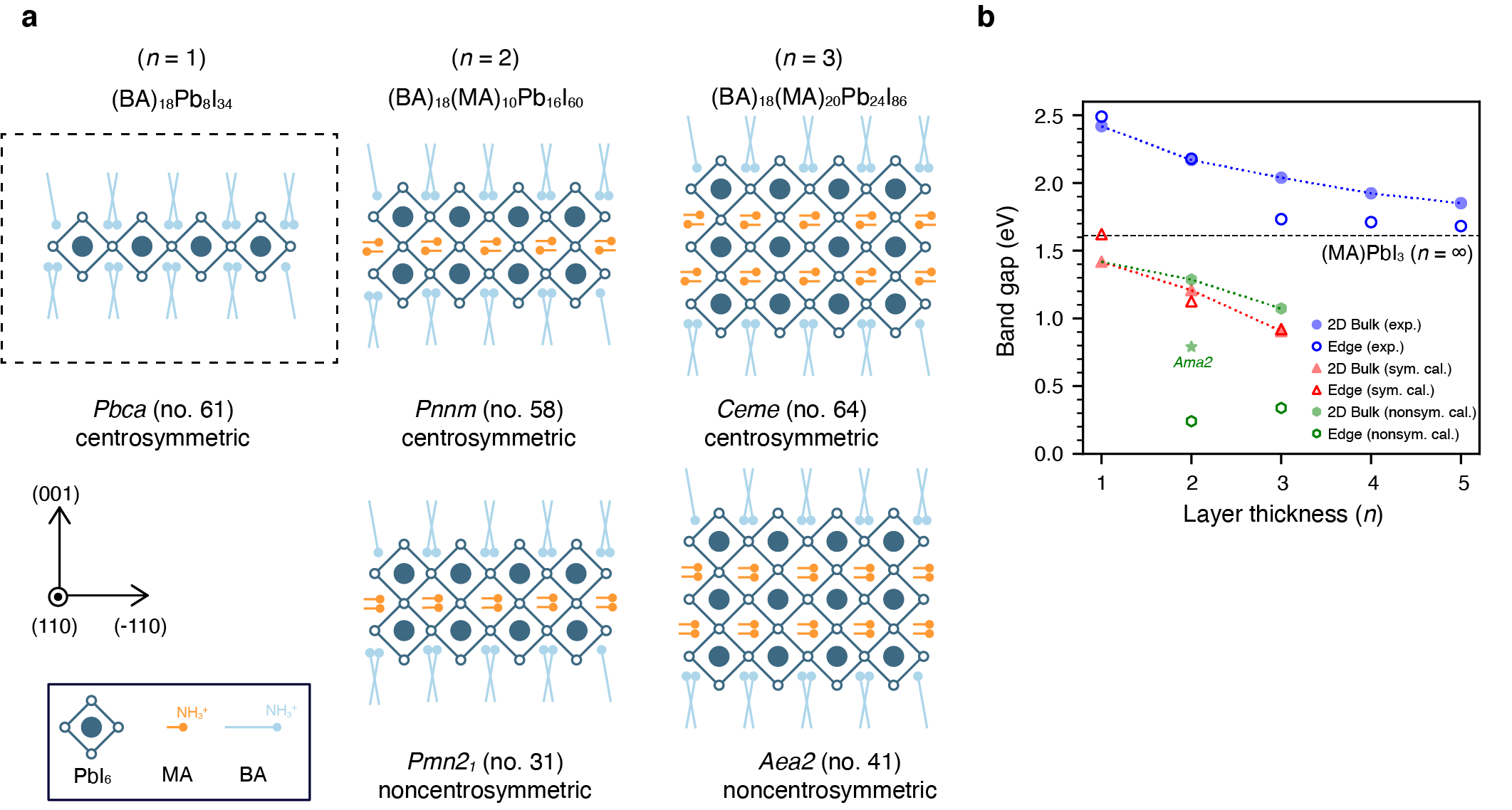}

\caption{\label{fig:structures}
(a) Schematics of ribbon structures of 2D perovskites used for edge electronic structure calculations, for $n=$ 1, 2, and 3 in centrosymmteric and noncentrosymmetric space groups. The dashed line is a guide for a unit cell. Here, ribbons of width $w=$ 4 were used in the calculations.  
(b) Band gaps of 2D perovskites in their interior (``2D bulk'') and at their edges from experiments \cite{Blancon17} and calculations. The black dashed line indicates the band gap of MAPbI$_3$ which is the 3D counterpart \cite{Yamada14}.}
\end{figure*}

Using DFT calculations, we have studied the structural and electronic properties of (BA)$_2$(MA)$_{n-1}$Pb$_n$I$_{3n+1}$ for $n=$ 1, 2, and 3 as a representative of 2D hybrid perovskites. Possible crystal structures consistent with X-ray diffraction include centrosymmetric as well as noncentrosymmetric variants \cite{Billing07,Stoumpos16}. In general, the noncentrosymmetric structural variants allow for ferroelectric ordering of the small organic MA cations situated between PbI$_6$ octahedra when $n\geq2$. To understand their relative stability, we have performed structural optimizations. As starting points, we included the configurations presented in Refs.~\citenum{Billing07,Stoumpos16}, and considered both centrosymmetric and noncentrosymmetic space groups for $n\geq2$. For the $n=$ 1 case, which has no orientable MA cations, we fully relaxed atomic positions of the structure in centrosymmeric $Pbca$ space group. For $n=$ 2 and 3, we find that the noncentrosymmetric structures in $Pmn2_1$ and $Aea2$ space groups are energetically more stable than their centrosymmetric counterparts, in $Pnnm$ and $Cmce$ space group, by 80.75 and 782 meV/f.u., respectively. Full structural details are reported in the Supporting Information.

In the centrosymmetric structures, neighboring MA cations point in the opposite directions and cancel the dipole moments of each other, whereas in the noncentrosymmetric structures, there are finite dipole moments. We note that in these preferred structures for $n=$ 2 and 3, the MAs lie in the direction parallel to the 2D perovskite layers, giving rise to in-plane internal electric fields, while the higher energy $Ama2$ structure for $n=$ 2 has MAs oriented perpendicular to the layers. Unlike the 3D perovskites, which have a flat energy surface containing many competing local minima for freely rotating MA ions \cite{Tan17,Lee16,Poglitsch87,Weller15,Leguy16}, the noncentrosymmetric structures found here are fairly stable by a few tens to hundreds meV/f.u. (see Supporting Information Table S1), compared to the centrosymmetric local energy minima. It is likely that large A cations play a role in restricting the fluctuations of the inorganic lattice \cite{Marchenko20}, and hence reducing the number of possible orientations of smaller A' cations. 

To understand the impact of A' cation ordering on layer edge electronic structure, we construct structural models of 2D perovskite layer edges for $n=$ 1, 2 and 3 layers, as ribbon structures in a supercell geometry, schematically shown in Figure~\ref{fig:structures}a. For each of the  2D bulk structures, we form edges along the (110) direction by cutting layers perpendicular to the (-110) direction. 
The ribbons considered here are four octahedral units wide. Charge neutrality is conserved in these calculations by removing one BA ion from each edge of the ribbon.

The calculated band gaps of 2D bulk and layer edges, and their dependence on layer thickness $n$ are summarized in Figure~\ref{fig:structures}b. For the 2D bulk, band gaps are expected to increase as $n$ decreases because of the increasing quantum confinement effect. This trend is captured by the DFT calculations, although the $Ama2$ 2D bulk structure ($n=$ 2) is an outlier in that it has a much smaller band gap (0.790 eV, green star) than the other 2D bulk phases. 
This occurs because a potential gradient perpendicular to the 2D perovskite layers is generated by vertically aligned MA ions in this phase (Supporting Information Figure S2), whereas the other phases have in-plane MA orientations. As a result, $n=$ 2 in the $Ama2$ space group does not follow the monotonically decreasing band gap trend experimentally observed \cite{Stoumpos16}. Because of this discrepancy, along with its higher formation energy than $Pmn2_1$ phase (845 meV/f.u.), we consider $Ama2$ to be an unlikely structure for $n=$ 2. Even though the overall values of the calculated band gaps are underestimated compared to experiments due to the underestimation of correlation effects in DFT, we expect that the trends arising from quantum confinement effects are accurately represented at the DFT level, as are the polarization induced electrostatic effects discussed below.

\begin{figure*}
\includegraphics{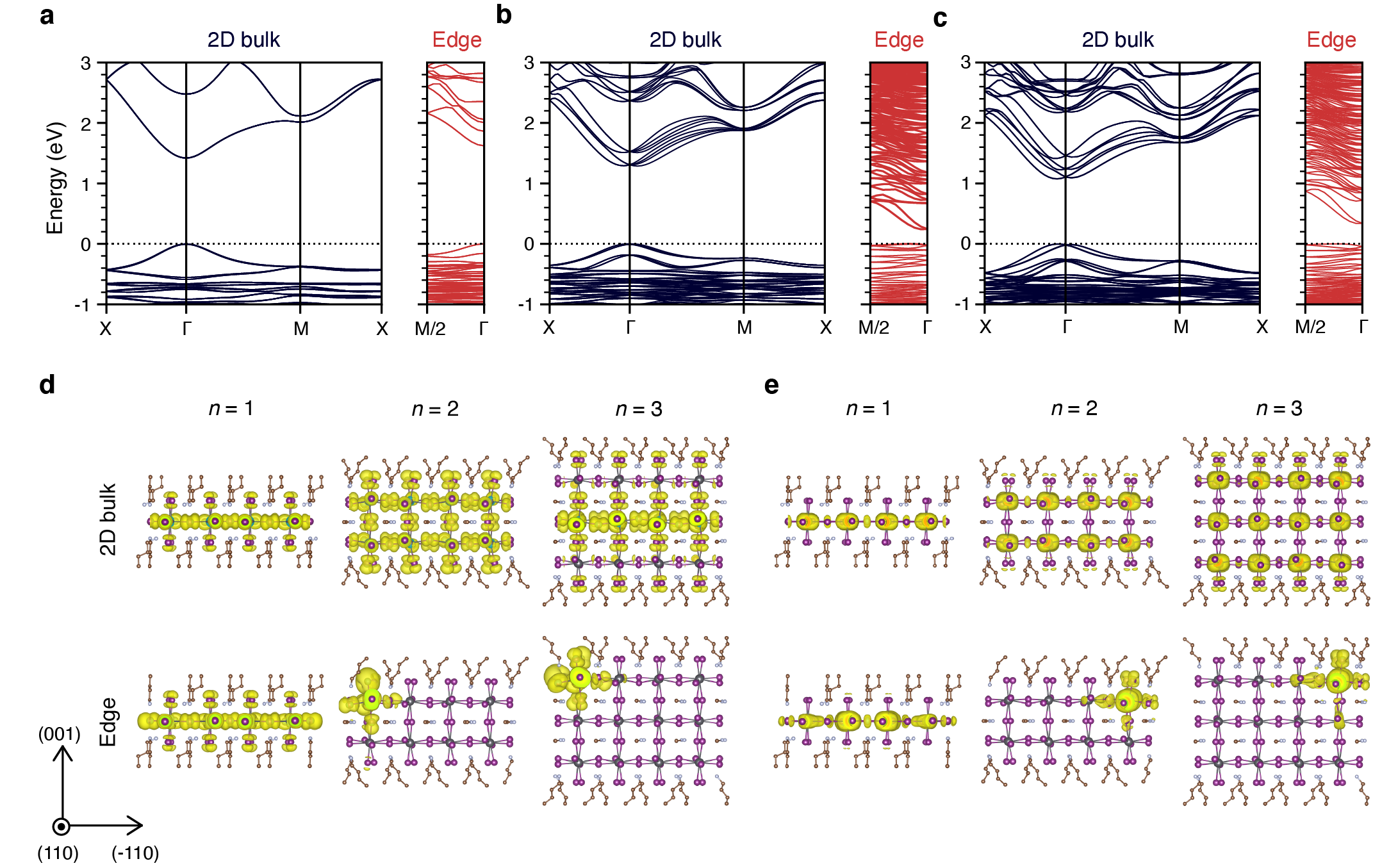}

\caption{\label{fig:edgestates}
Band structures of 2D bulk and edges of perovskites for (a) $n=$ 1, (b) $n=$ 2, and (c) $n=$ 3 are shown. The band structures are plotted along k-paths X(1/2, 0, 0)--$\Gamma$(0, 0, 0)--M(1/2, 1/2, 0)--X(1/2, 0, 0) and M/2(1/2, 0, 0)--$\Gamma$(0, 0, 0) for bulk and ribbon, respectively. 
The partial charge densities of (d) the valence band maximum and (e) the conduction band minimum of 2D bulk and ribbon perovskites at $\Gamma$-point are shown for $n=$ 1, 2, and 3. These are illustrated in the same orientation as Figure~\ref{fig:structures}a. }
\end{figure*}

Next, we compare the band gaps of 2D bulk and edges. 
As the system is confined further from a 2D layer to a ribbon, additional quantum confinement effects arise which perturb the electronic structure towards increasing band gaps. This effect is the strongest in $n=$ 1, in the calculations as well as in the experiments, while the centrosymmetric structures for $n\geq2$ do not show much change in their band gaps between 2D bulk and ribbon forms. On the other hand, other factors arise as layer edges are formed from noncentrosymmetric 2D bulk structures. As we show below, the presence of in-plane polarization strongly modifies the edge electronic structure in these cases. They show a large decrease in band gap when the 2D bulk structures are compared to ribbons, by 1.04 eV and 0.735 eV for $n=$ 2 and 3, respectively.
This suggests that lower energy states at the edges are linked to symmetry breaking in 2D perovskites. 
Bulk and edge band structures in Figures~\ref{fig:edgestates}a--c show that this band gap narrowing arises from a shift in valence and conduction bands, instead of the formation of mid-gap states. 
Incidentally, Rashba spin splittings around the $\Gamma$-point occur for the noncentrosymmetric structures when $n>$ 2 but not for $n=$ 1, due to the requirement of broken inversion symmetry \cite{Zhai17}. 
The band gap drops expected by the simulations are larger than those observed in experiments, because MAs fluctuate around their average symmetric orientations at room temperature \cite{Tan17,Lee16,Poglitsch87,Weller15,Leguy16}, which is not considered in our calculations.
In addition, while the edge states have been reported only for (BA)$_2$(MA)$_{n-1}$Pb$_n$I$_{3n+1}$ with $n\geq3$, our calculations expect the lower energy edge states even for $n=2$. Although it is expected the noncentrosymmetric phases to be energetically favorable for both $n=2$ and $n=3$ from our DFT calculations, where the consideration for entropy and thermodynamic effects at finite temperature are missing, actual perovskite samples could exist in different phases at room temperature.
The energy difference between centrosymmetric and noncentrosymmetric phases is about ten times shallower for $n=2$ than $n=3$, and $n=2$ could be exist in centrosymmetric form at room temperature, showing no evidence of edge states. 

Partial charge densities of the valence band maximum (VBM) and conduction band minimum (CBM) of 2D bulk and edges at the $\Gamma$-point are presented in Figures~\ref{fig:edgestates}d and e, showing that formation of lower-energy localized states at the edges is responsible for band gap narrowing. 
For $n=$ 1, the partial charge densities of 2D ribbon are dispersed over the entire layer, just like those of their 2D bulk parent structure. 
On the other hand, for $n\geq$ 2, the wave functions of the VBM and CBM are strongly localized at the edges.
These DFT calculations confirm that strongly localized edge states form in noncentrosymmetric 2D perovskites where MA cations are aligned with each other and consequent internal electric fields are, explaining why edge states are only observed in experiments when $n>1$ \cite{Blancon17,Shi19,Zhao19}.

\begin{figure*}
\includegraphics{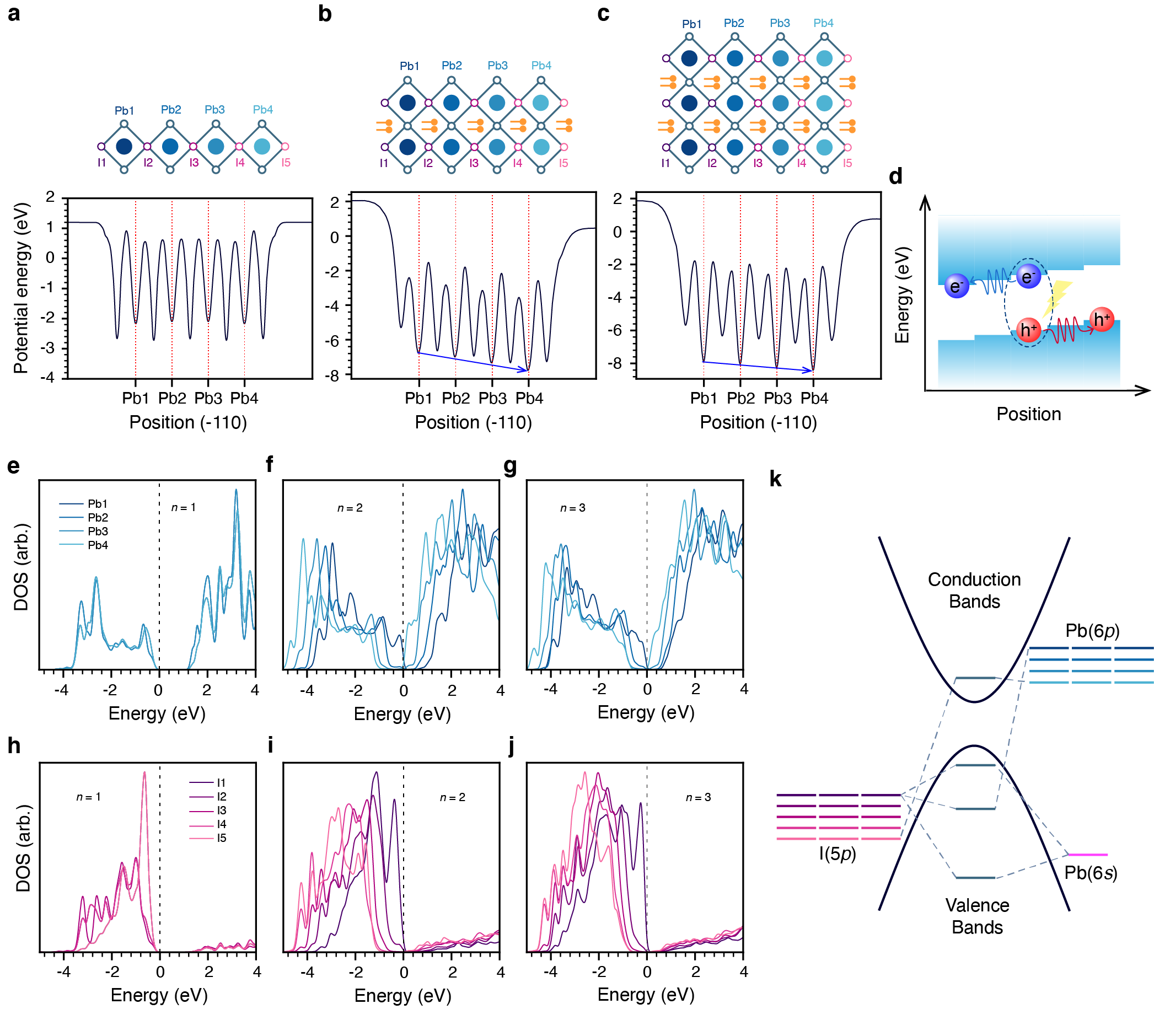}

\caption{\label{fig:mechanism}
Plane-averaged local potentials of 2D perovskite ribbons along (110) for (a) $n=$ 1, (b) $n=$ 2, and (c) $n=$ 3. 
(d) A schematic diagram of the mechanism of formation of the localized edge states with lower energies are presented.
Projected density of states of Pb and I atoms are plotted for (e, h) $n=$ 1, (f, i) $n=$ 2, and (g, j) $n=$ 3. 
(k) The orbital diagram of a 2D perovskite ribbon structure with finite dipole moment across it is shown.
}
\end{figure*}

The origin of layer edge states in the noncentrosymmetric structures is the formation of in-plane internal electric  fields. 
Plane-averaged local potentials along the (-110) direction plotted in Figures~\ref{fig:mechanism}a--c show that a potential gradient is created by aligned MA dipole moments in noncentrosymmetric configurations. The perovskite ribbons with $n>$ 1 have finite dipole moments of 4.237 e$\cdot$\AA{} and 3.747 e$\cdot$\AA{} along the (-110) direction for $n=$ 2 and 3, respectively.
Because of this, atomic sites located at different distances away from the edge experience different electrostatic potentials, as seen in the projected density of states (PDOS) plotted in Figures~\ref{fig:mechanism}e--j. 
For $n=$ 1, the PDOS of each Pb and I atoms are independent of distance from the edge.
However, for $n>$ 1, the PDOS of Pb and I show clear energy shifts from the higher to lower local potential regions.
In these halide perovskites, the electronic states near the Fermi level are predominantly of I $5p$ and Pb $6p$ orbital character. 
The energy levels of these atomic orbitals are split under the internal electric field, leading to a smaller band gap between the band edge states that are naturally located at opposite edges (Figure~\ref{fig:mechanism}k).
In our calculations, the edge states are localized at the top corners of the edges instead of being spread over the entire thickness of the edge in the (001) direction (Figures~\ref{fig:edgestates}d and e). This is because the absence of BA ions at one of the corners additionally breaks the symmetry between the top and bottom corners of each edge. In actual samples, we expect substrate effects and the details of ligand interactions to determine the exact position of the states on the edge \cite{Kilina09}.

While our DFT layer edge state calculations are restricted to 2D ribbons of finite width, the conclusions extend to semi-infinite layer edges of macroscopic 2D perovskite flakes. To show this, we have applied a simple tight binding model (see Supporting Information for details) to calculate layer edge states of large width ribbons. We find that,  as in the DFT simulations, edge states in the large-width tight-binding model form when there is an in-plane electric field normal to the edge, and not without such an electric field. It shows that electronic polarization plays an important role in stabilizing layer edge states. This result holds for all layer thicknesses (cases $n=$ 1, 2, 3 are explicitly illustrated) as long as this electric field can be supported.   

\begin{figure}
\includegraphics{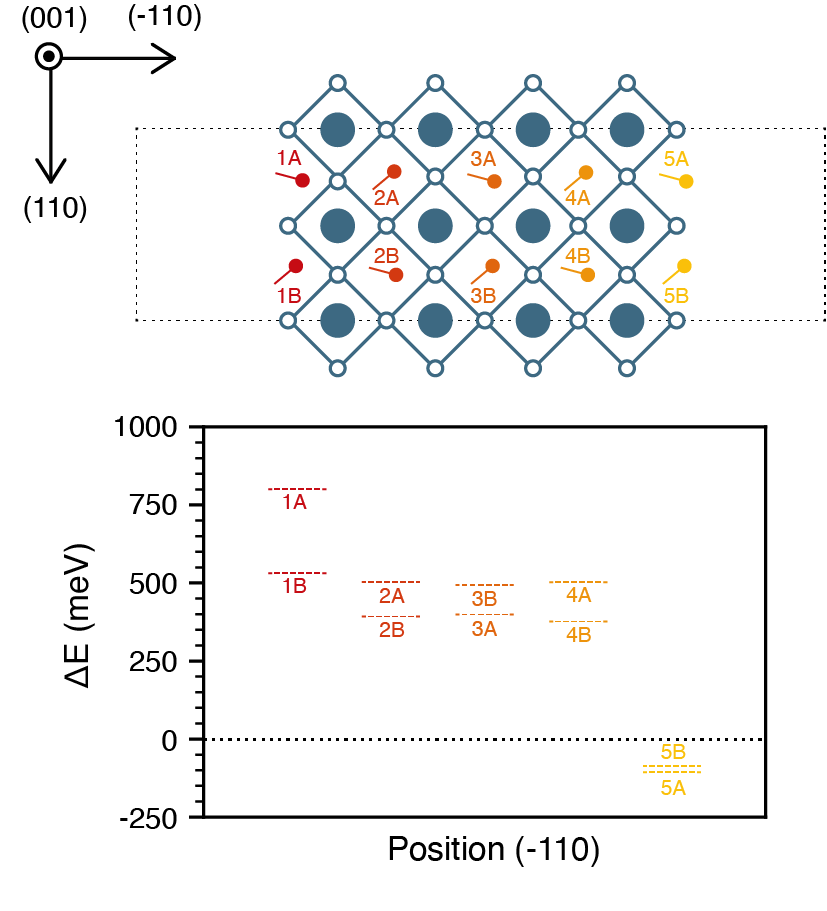}

\caption{\label{fig:ma_flip}
Top view of $n=$ 2 ribbon structure unit cell and the total energy gain or loss when a MA ion is reversed. The BA ions are omitted in the structure for clear view and the total energy of the ribbon without flipping a MA ion is used as a reference level.
}
\end{figure}

Besides serving as a location for electronic localization, the edges of 2D perovskites also play a structural role in determining the direction of ferroelectric alignment of MA cations. It is known that the NH$_3^{+}$ group of MA preferentially forms hydrogen bonds with the halide anions in halide perovskites \cite{Tan17,Y.Yang18}. This fixes the direction of the outermost MA cations at the edges (Figure~\ref{fig:ma_flip}, red), creating a polarization field that aligns other MA cations adjacent to it and away from the edges.  
To investigate the stability of this process, we calculated the total energy gain or loss from flipping each MA ion in a $n=$ 2 ribbon structure as shown in Figure~\ref{fig:ma_flip}.
At the edge with preferential NH$_3^{+}$ hydrogen bonding, it requires more than 500 meV to flip MA cations (1A or 1B), whereas CH$_3$ hydrogen bonded edges (5A or 5B) are unstable and release energy of about 100 meV upon flipping of MA cation directions.
MA ions near the edge (2A, 2B,..., 4B) tend to align in the same direction as the outermost MAs, having positive flipping energies of several hundred meV.  This propagates at least up to the fourth layer into the edge, which is the size of the simulation. Additionally, our $ab-initio$ MD (AIMD) simulations of bulk 2D perovskites suggest that some ferroelectric polarization effects persist even in the absence of edge effects, and at room temperature. These trajectories show that the $n=$ 3 structure in particular displays a tendency for stable in-plane MA ordering even without anchoring at the edges (Supporting Information Figure S3).

Internal electric fields generated at the edges this way are expected to aid carrier separation. Electrons will preferentially occupy edge regions because of the direction of the electric field set by NH$_3^{+}$ hydrogen bonding at the edges, while holes will disperse into the bulk region. Exciton dissociation and separation of electrons and holes by this mechanism would result in longer luminescence lifetimes at the edges than in the bulk, explaining the observations in Refs.~\citenum{Blancon17,Shi19,Zhao19}. 
Other electron-hole separation mechanisms can possibly coexist at the edge. 
Local optical and electronic experimental probes, such as scanning probe microscopies, would provide interesting directions for further understanding of the edge electronic structure in these materials.

In conclusion, we have uncovered an intrinsic driver for low-energy localized edge state formation in 2D hybrid perovskites using first-principles calculations.
Our work shows that layer edge states are stabilized by the ferroelectric alignment of organic cations, leading to charge separation and long carrier lifetimes at 2D hybrid perovskites edges. 
The mechanism presented here enables further design and manipulation of exciton dissociation pathways in 2D hybrid perovksites.

\section{Methods}

For DFT calculations, we used the projector-augmented-plane-wave method as implemented in the Vienna \textit{Ab-initio} Simulation Package (VASP) \cite{Kresse96, Kresse99}. We used experimentally reported cell parameters of (BA)$_2$(MA)$_{n-1}$Pb$_{n}$I$_{3n+1}$ from Refs.~\citenum{Billing07, Stoumpos16} measured by single crystal X-ray diffraction at room temperature for $n=$1, 2 and 3. 
We fully relaxed atomic positions until the force between atoms became less than 0.01 eV/\AA{} using a 500 eV plane-wave cutoff and a $5\times5\times1$ Monkhorst-Pack\cite{Monkhorst76} k-point mesh. 
We chose to use the optB88-vdW functional \cite{Klimes10, Klimes11} as it showed the best agreement with experiment in crystallographic parameters of 3D hybrid perovskite (MA)PbI$_3$ \cite{Menendez-Proupin14}. 

To investigate electronic structures and total energy changes as MAs flip, we used the PBE-GGA functional \cite{Perdew96} with convergence criterion of 10$^{-6}$ eV. 
For 2D bulk, a $5\times5\times1$ Monkhorst-Pack k-point mesh was used and for ribbons, extended along the Cartesian $x$-axis, $3\times1\times1$ was used.
Dipole correction was considered for noncentrosymmetric ribbon structures. 
Spin-orbit coupling was included for electronic structure calculations but not for structural relaxation and total energy calculation.
Further details are shown in Supporting Information.

\begin{acknowledgement}
 We are grateful to Alex Weber-Bargioni, Brett A. Helms, Kevin Whitham, 
 Monica Lorenzon, and Ruoxi Yang for fruitful discussions.
 The authors were supported by the Molecular Foundry, a DOE Office of Science User Facility supported by the Office of Science of the U.S. Department of Energy under Contract No. DE-AC02-05CH11231. This research used resources of the National Energy Research Scientific Computing Center, a DOE Office of Science User Facility supported by the Office of Science of the U.S. Department of Energy under Contract No. DE-AC02-05CH11231.  

\end{acknowledgement}



\bibliography{manuscript}

\providecommand{\latin}[1]{#1}
\makeatletter
\providecommand{\doi}
  {\begingroup\let\do\@makeother\dospecials
  \catcode`\{=1 \catcode`\}=2 \doi@aux}
\providecommand{\doi@aux}[1]{\endgroup\texttt{#1}}
\makeatother
\providecommand*\mcitethebibliography{\thebibliography}
\csname @ifundefined\endcsname{endmcitethebibliography}
  {\let\endmcitethebibliography\endthebibliography}{}
\begin{mcitethebibliography}{42}
\providecommand*\natexlab[1]{#1}
\providecommand*\mciteSetBstSublistMode[1]{}
\providecommand*\mciteSetBstMaxWidthForm[2]{}
\providecommand*\mciteBstWouldAddEndPuncttrue
  {\def\EndOfBibitem{\unskip.}}
\providecommand*\mciteBstWouldAddEndPunctfalse
  {\let\EndOfBibitem\relax}
\providecommand*\mciteSetBstMidEndSepPunct[3]{}
\providecommand*\mciteSetBstSublistLabelBeginEnd[3]{}
\providecommand*\EndOfBibitem{}
\mciteSetBstSublistMode{f}
\mciteSetBstMaxWidthForm{subitem}{(\alph{mcitesubitemcount})}
\mciteSetBstSublistLabelBeginEnd
  {\mcitemaxwidthsubitemform\space}
  {\relax}
  {\relax}

\bibitem[Wolf \latin{et~al.}(2014)Wolf, Holovsky, Moon, L{\"o}per, Niesen,
  Ledinsky, Haug, Yum, and Ballif]{Wolf14}
Wolf,~S.~D.; Holovsky,~J.; Moon,~S.-J.; L{\"o}per,~P.; Niesen,~B.;
  Ledinsky,~M.; Haug,~F.-J.; Yum,~J.-H.; Ballif,~C. {Organometallic Halide
  Perovskites: Sharp Optical Absorption Edge and Its Relation to Photovoltaic
  Performance}. \emph{J. Phys. Chem. Lett.} \textbf{2014}, \emph{5},
  1035--1039\relax
\mciteBstWouldAddEndPuncttrue
\mciteSetBstMidEndSepPunct{\mcitedefaultmidpunct}
{\mcitedefaultendpunct}{\mcitedefaultseppunct}\relax
\EndOfBibitem
\bibitem[Oga \latin{et~al.}(2014)Oga, Saeki, Ogomi, Hayase, and Seki]{Oga14}
Oga,~H.; Saeki,~A.; Ogomi,~Y.; Hayase,~S.; Seki,~S. {Improved Understanding of
  the Electronic and Energetic Landscapes of Perovskite Solar Cells: High Local
  Charge Carrier Mobility, Reduced Recombination, and Extremely Shallow Traps}.
  \emph{J. Am. Chem. Soc.} \textbf{2014}, \emph{136}, 13818--13825\relax
\mciteBstWouldAddEndPuncttrue
\mciteSetBstMidEndSepPunct{\mcitedefaultmidpunct}
{\mcitedefaultendpunct}{\mcitedefaultseppunct}\relax
\EndOfBibitem
\bibitem[Stranks \latin{et~al.}(2013)Stranks, Eperon, Grancini, Menelaou,
  Alcocer, Leijtens, Herz, Petrozza, and Snaith]{Stranks13}
Stranks,~S.~D.; Eperon,~G.~E.; Grancini,~G.; Menelaou,~C.; Alcocer,~M. J.~P.;
  Leijtens,~T.; Herz,~L.~M.; Petrozza,~A.; Snaith,~H.~J. {Electron-Hole
  Diffusion Lengths Exceeding 1 Micrometer in an Organometal Trihalide
  Perovskite Absorber}. \emph{Science} \textbf{2013}, \emph{342},
  341--344\relax
\mciteBstWouldAddEndPuncttrue
\mciteSetBstMidEndSepPunct{\mcitedefaultmidpunct}
{\mcitedefaultendpunct}{\mcitedefaultseppunct}\relax
\EndOfBibitem
\bibitem[Xing \latin{et~al.}(2013)Xing, Mathews, Sun, Lim, Lam, Gr{\"a}tzel,
  Mhaisalkar, and Sum]{Xing13}
Xing,~G.; Mathews,~N.; Sun,~S.; Lim,~S.~S.; Lam,~Y.~M.; Gr{\"a}tzel,~M.;
  Mhaisalkar,~S.; Sum,~T.~C. {Long-Range Balanced Electron- and Hole-Transport
  Lengths in Organic-Inorganic CH$_3$NH$_3$PbI$_3$}. \emph{Science}
  \textbf{2013}, \emph{342}, 344--347\relax
\mciteBstWouldAddEndPuncttrue
\mciteSetBstMidEndSepPunct{\mcitedefaultmidpunct}
{\mcitedefaultendpunct}{\mcitedefaultseppunct}\relax
\EndOfBibitem
\bibitem[Shi \latin{et~al.}(2015)Shi, Adinolfi, Comin, Yuan, Alarousu, Buin,
  Chen, Hoogland, Rothenberger, Katsiev, Losovyj, Zhang, Dowben, Mohammed,
  Sargent, and Bakr]{Shi15}
Shi,~D. \latin{et~al.}  {Low trap-state density and long carrier diffusion in
  organolead trihalide perovskite single crystals}. \emph{Science}
  \textbf{2015}, \emph{347}, 519--522\relax
\mciteBstWouldAddEndPuncttrue
\mciteSetBstMidEndSepPunct{\mcitedefaultmidpunct}
{\mcitedefaultendpunct}{\mcitedefaultseppunct}\relax
\EndOfBibitem
\bibitem[Dong \latin{et~al.}(2015)Dong, Fang, Shao, Mulligan, Qiu, Cao, and
  Huang]{Dong15}
Dong,~Q.; Fang,~Y.; Shao,~Y.; Mulligan,~P.; Qiu,~J.; Cao,~L.; Huang,~J.
  {Electron-hole diffusion lengths $>$ 175 $\mu$m in solution-grown
  CH$_3$NH$_3$PbI$_3$ single crystals}. \emph{Science} \textbf{2015},
  \emph{347}, 967--970\relax
\mciteBstWouldAddEndPuncttrue
\mciteSetBstMidEndSepPunct{\mcitedefaultmidpunct}
{\mcitedefaultendpunct}{\mcitedefaultseppunct}\relax
\EndOfBibitem
\bibitem[Kojima \latin{et~al.}(2009)Kojima, Teshima, Shirai, and
  Miyasaka]{Kojima09}
Kojima,~A.; Teshima,~K.; Shirai,~Y.; Miyasaka,~T. {Organometal Halide
  Perovskites as Visible-Light Sensitizers for Photovoltaic Cells}. \emph{J.
  Am. Chem. Soc.} \textbf{2009}, \emph{131}, 6050--6051\relax
\mciteBstWouldAddEndPuncttrue
\mciteSetBstMidEndSepPunct{\mcitedefaultmidpunct}
{\mcitedefaultendpunct}{\mcitedefaultseppunct}\relax
\EndOfBibitem
\bibitem[NREL()]{NREL}
NREL, Best Research-Cell Efficiencies chart.
  \url{https://www.nrel.gov/pv/assets/pdfs/best-research-cell-efficiencies.20200406.pdf}\relax
\mciteBstWouldAddEndPuncttrue
\mciteSetBstMidEndSepPunct{\mcitedefaultmidpunct}
{\mcitedefaultendpunct}{\mcitedefaultseppunct}\relax
\EndOfBibitem
\bibitem[Smith \latin{et~al.}(2014)Smith, Hoke, Solis‐Ibarra, McGehee, and
  Karunadasa]{Smith14}
Smith,~I.~C.; Hoke,~E.~T.; Solis‐Ibarra,~D.; McGehee,~M.~D.;
  Karunadasa,~H.~I. {A Layered Hybrid Perovskite Solar-Cell Absorber with
  Enhance Moisture Stability}. \emph{Angew. Chem. Int. Ed.} \textbf{2014},
  \emph{53}, 11232--11235\relax
\mciteBstWouldAddEndPuncttrue
\mciteSetBstMidEndSepPunct{\mcitedefaultmidpunct}
{\mcitedefaultendpunct}{\mcitedefaultseppunct}\relax
\EndOfBibitem
\bibitem[Cao \latin{et~al.}(2015)Cao, Stoumpos, Farha, Hupp, and
  Kanatzidis]{Cao15}
Cao,~D.~H.; Stoumpos,~C.~C.; Farha,~O.~K.; Hupp,~J.~T.; Kanatzidis,~M.~G. {2D
  Homologous Perovskites as Light-Absorbing Materials for Solar Cell
  Applications}. \emph{J. Am. Chem. Soc.} \textbf{2015}, \emph{137},
  7843--7850\relax
\mciteBstWouldAddEndPuncttrue
\mciteSetBstMidEndSepPunct{\mcitedefaultmidpunct}
{\mcitedefaultendpunct}{\mcitedefaultseppunct}\relax
\EndOfBibitem
\bibitem[Tsai \latin{et~al.}(2016)Tsai, Nie, Blancon, Stoumpos, Asadpour,
  Harutyunyan, Neukirch, Verduzco, Crochet, Tretiak, Pedesseau, Even, Alam,
  Gupta, Lou, Ajayan, Bedzyk, Kanatzidis, and Mohite]{Tsai16}
Tsai,~H. \latin{et~al.}  {High-efficiency two-dimensional Ruddlesden-Popper
  perovskite solar cells}. \emph{Nature} \textbf{2016}, \emph{536},
  312--316\relax
\mciteBstWouldAddEndPuncttrue
\mciteSetBstMidEndSepPunct{\mcitedefaultmidpunct}
{\mcitedefaultendpunct}{\mcitedefaultseppunct}\relax
\EndOfBibitem
\bibitem[Tsai \latin{et~al.}(2018)Tsai, Nie, Blancon, Stoumpos, Soe, Yoo,
  Tretiak, Even, Sadhanala, Azzellino, Brenes, Ajayan, Bulovi{\'c}, Stranks,
  Friend, Kanatzidis, and Mohite]{Tsai18}
Tsai,~H. \latin{et~al.}  {Stable Light‐Emitting Diodes Using Phase‐Pure
  Ruddlesden–Popper Layered Perovskites}. \emph{Adv. Mater.} \textbf{2018},
  \emph{30}, 1704217\relax
\mciteBstWouldAddEndPuncttrue
\mciteSetBstMidEndSepPunct{\mcitedefaultmidpunct}
{\mcitedefaultendpunct}{\mcitedefaultseppunct}\relax
\EndOfBibitem
\bibitem[Yang \latin{et~al.}(2018)Yang, Zhang, Deng, Chu, Jiang, Meng, Wang,
  Zhang, Yin, and You]{Yang18}
Yang,~X.; Zhang,~X.; Deng,~J.; Chu,~Z.; Jiang,~Q.; Meng,~J.; Wang,~P.;
  Zhang,~L.; Yin,~Z.; You,~J. {Efficient green light-emitting diodes based on
  quasi-two-dimensional composition and phase engineered perovskite with
  surface passivation}. \emph{Nat. Commun.} \textbf{2018}, \emph{9}, 570\relax
\mciteBstWouldAddEndPuncttrue
\mciteSetBstMidEndSepPunct{\mcitedefaultmidpunct}
{\mcitedefaultendpunct}{\mcitedefaultseppunct}\relax
\EndOfBibitem
\bibitem[Billing and Lemmerer(2007)Billing, and Lemmerer]{Billing07}
Billing,~D.; Lemmerer,~A. {Synthesis, characterization and phase transitions in
  the inorganic-organic layered perovskite-type hybrids
  [(C$_n$H$_{2n+1}$NH$_3$)$_2$PbI$_4$], n = 4, 5 and 6}. \emph{Acta
  Crystallogr. B} \textbf{2007}, \emph{63}, 735--747\relax
\mciteBstWouldAddEndPuncttrue
\mciteSetBstMidEndSepPunct{\mcitedefaultmidpunct}
{\mcitedefaultendpunct}{\mcitedefaultseppunct}\relax
\EndOfBibitem
\bibitem[Stoumpos \latin{et~al.}(2016)Stoumpos, Cao, Clark, Young, Rondinelli,
  Jang, Hupp, and Kanatzidis]{Stoumpos16}
Stoumpos,~C.~C.; Cao,~D.~H.; Clark,~D.~J.; Young,~J.; Rondinelli,~J.~M.;
  Jang,~J.~I.; Hupp,~J.~T.; Kanatzidis,~M.~G. {Ruddlesden-Popper Hybrid Lead
  Iodide Perovskite 2D Homologous Semiconductors}. \emph{Chem. Mater.}
  \textbf{2016}, \emph{28}, 2852--2867\relax
\mciteBstWouldAddEndPuncttrue
\mciteSetBstMidEndSepPunct{\mcitedefaultmidpunct}
{\mcitedefaultendpunct}{\mcitedefaultseppunct}\relax
\EndOfBibitem
\bibitem[Soe \latin{et~al.}(2019)Soe, Nagabhushana, Shivaramaiah, Tsai, Nie,
  Blancon, Melkonyan, Cao, Traor{\'e}, Pedesseau, Kepenekian, Katan, Even,
  Marks, Navrotsky, Mohite, Stoumpos, and Kanatzidis]{soe19}
Soe,~C. M.~M. \latin{et~al.}  Structural and thermodynamic limits of layer
  thickness in 2D halide perovskites. \emph{Proc. Natl. Acad. Sci. U.S.A.}
  \textbf{2019}, \emph{116}, 58--66\relax
\mciteBstWouldAddEndPuncttrue
\mciteSetBstMidEndSepPunct{\mcitedefaultmidpunct}
{\mcitedefaultendpunct}{\mcitedefaultseppunct}\relax
\EndOfBibitem
\bibitem[Yang \latin{et~al.}(2018)Yang, Gao, Gao, and Wei]{Y.Yang18}
Yang,~Y.; Gao,~F.; Gao,~S.; Wei,~S.-H. {Origin of the stability of
  two-dimensional perovskites: a first-principles study}. \emph{J. Mater. Chem.
  A} \textbf{2018}, \emph{6}, 14949--14955\relax
\mciteBstWouldAddEndPuncttrue
\mciteSetBstMidEndSepPunct{\mcitedefaultmidpunct}
{\mcitedefaultendpunct}{\mcitedefaultseppunct}\relax
\EndOfBibitem
\bibitem[Umebayashi \latin{et~al.}(2003)Umebayashi, Asai, Kondo, and
  Nakao]{Umebayashi03}
Umebayashi,~T.; Asai,~K.; Kondo,~T.; Nakao,~A. {Electronic structures of lead
  iodide based low-dimensional crystals}. \emph{Phys. Rev. B} \textbf{2003},
  \emph{67}, 155405\relax
\mciteBstWouldAddEndPuncttrue
\mciteSetBstMidEndSepPunct{\mcitedefaultmidpunct}
{\mcitedefaultendpunct}{\mcitedefaultseppunct}\relax
\EndOfBibitem
\bibitem[Blancon \latin{et~al.}(2017)Blancon, Tsai, Nie, Stoumpos, Pedesseau,
  Katan, Kepenekian, Soe, Appavoo, Sfeir, Tretiak, Ajayan, Kanatzidis, Even,
  Crochet, and Mohite]{Blancon17}
Blancon,~J.-C. \latin{et~al.}  {Extremely efficient internal exciton
  dissociation through edge states in layered 2D perovskites}. \emph{Science}
  \textbf{2017}, \emph{355}, 1288--1292\relax
\mciteBstWouldAddEndPuncttrue
\mciteSetBstMidEndSepPunct{\mcitedefaultmidpunct}
{\mcitedefaultendpunct}{\mcitedefaultseppunct}\relax
\EndOfBibitem
\bibitem[Shi \latin{et~al.}(2019)Shi, Deng, Yuan, Gao, Akriti, Yuan, Davis,
  Zemlyanov, Yu, Huang, and Dou]{Shi19}
Shi,~E.; Deng,~S.; Yuan,~B.; Gao,~Y.; Akriti,; Yuan,~L.; Davis,~C.~S.;
  Zemlyanov,~D.; Yu,~Y.; Huang,~L.; Dou,~L. {Extrinsic and Dynamic Edge States
  of Two-Dimensional Lead Halide Perovskites}. \emph{ACS Nano} \textbf{2019},
  \emph{13}, 1635--1644\relax
\mciteBstWouldAddEndPuncttrue
\mciteSetBstMidEndSepPunct{\mcitedefaultmidpunct}
{\mcitedefaultendpunct}{\mcitedefaultseppunct}\relax
\EndOfBibitem
\bibitem[Zhao \latin{et~al.}(2019)Zhao, Tian, Leng, Zhao, and Jin]{Zhao19}
Zhao,~C.; Tian,~W.; Leng,~J.; Zhao,~Y.; Jin,~S. {Controlling the Property of
  Edges in Layered 2D Perovskite Single Crystals}. \emph{J. Phys. Chem. Lett.}
  \textbf{2019}, \emph{10}, 3950--3954\relax
\mciteBstWouldAddEndPuncttrue
\mciteSetBstMidEndSepPunct{\mcitedefaultmidpunct}
{\mcitedefaultendpunct}{\mcitedefaultseppunct}\relax
\EndOfBibitem
\bibitem[Feng \latin{et~al.}(2018)Feng, Gong, Gao, Wen, Gong, Jiang, Zhang, Wu,
  Wu, Fu, Jiang, and Zhang]{Feng18}
Feng,~J.; Gong,~C.; Gao,~H.; Wen,~W.; Gong,~Y.; Jiang,~X.; Zhang,~B.; Wu,~Y.;
  Wu,~Y.; Fu,~H.; Jiang,~L.; Zhang,~X. {Single-crystalline layered metal-halide
  perovskite nanowires for ultrasensitive photodetectors}. \emph{Nat.
  Electron.} \textbf{2018}, \emph{1}, 404--410\relax
\mciteBstWouldAddEndPuncttrue
\mciteSetBstMidEndSepPunct{\mcitedefaultmidpunct}
{\mcitedefaultendpunct}{\mcitedefaultseppunct}\relax
\EndOfBibitem
\bibitem[Kepenekian \latin{et~al.}(2018)Kepenekian, Traore, Blancon, Pedesseau,
  Tsai, Nie, Stoumpos, Kanatzidis, Even, Mohite, Treiak, and
  Katan]{Kepenekian18}
Kepenekian,~M.; Traore,~B.; Blancon,~J.-C.; Pedesseau,~L.; Tsai,~H.; Nie,~W.;
  Stoumpos,~C.~C.; Kanatzidis,~M.~G.; Even,~J.; Mohite,~A.~D.; Treiak,~S.;
  Katan,~C. {Concept of Lattice Mismath and Emergence of Surface States in
  Two-dimensional Hybrid Perovskite Quantum Wells}. \emph{Nano Lett.}
  \textbf{2018}, \emph{18}, 5603--5609\relax
\mciteBstWouldAddEndPuncttrue
\mciteSetBstMidEndSepPunct{\mcitedefaultmidpunct}
{\mcitedefaultendpunct}{\mcitedefaultseppunct}\relax
\EndOfBibitem
\bibitem[Zhang \latin{et~al.}(2019)Zhang, Fang, Long, and Prezhdo]{Zhang19}
Zhang,~Z.; Fang,~W.-H.; Long,~R.; Prezhdo,~O.~V. {Exciton Dissociation and
  Suppressed Charge Recombination at 2D Perovskite Edges: Key Roles of
  Unsaturated Halide Bonds and Thermal Disorder}. \emph{J. Am. Chem. Soc.}
  \textbf{2019}, \emph{141}, 15557--15566\relax
\mciteBstWouldAddEndPuncttrue
\mciteSetBstMidEndSepPunct{\mcitedefaultmidpunct}
{\mcitedefaultendpunct}{\mcitedefaultseppunct}\relax
\EndOfBibitem
\bibitem[Qin \latin{et~al.}(2020)Qin, Dai, Gajjela, Wang, Hadjiev, Yang, Li,
  Zhong, Tang, Yao, Guloy, Reddy, Mayerich, Deng, Yu, Feng, Calderon,
  Hernandez, Wang, and Bao]{Qin20}
Qin,~Z. \latin{et~al.}  {Spontaneous Formation of 2D/3D Heterostructures on the
  Edges of 2D Ruddlesden–Popper Hybrid Perovskite Crystals}. \emph{Chem.
  Mater.} \textbf{2020}, \emph{32}, 5009--–5015\relax
\mciteBstWouldAddEndPuncttrue
\mciteSetBstMidEndSepPunct{\mcitedefaultmidpunct}
{\mcitedefaultendpunct}{\mcitedefaultseppunct}\relax
\EndOfBibitem
\bibitem[Yamada \latin{et~al.}(2014)Yamada, Nakamura, Endo, Wakamiya, and
  Kanemitsu]{Yamada14}
Yamada,~Y.; Nakamura,~T.; Endo,~M.; Wakamiya,~A.; Kanemitsu,~Y. {Near-band-edge
  optical responses of solution-processed organic–inorganic hybrid perovskite
  CH$_3$NH$_3$PbI$_3$ on mesoporous TiO$_2$ electrodes}. \emph{Appl. Phys.
  Express} \textbf{2014}, \emph{7}, 032302\relax
\mciteBstWouldAddEndPuncttrue
\mciteSetBstMidEndSepPunct{\mcitedefaultmidpunct}
{\mcitedefaultendpunct}{\mcitedefaultseppunct}\relax
\EndOfBibitem
\bibitem[Tan \latin{et~al.}(2017)Tan, Zheng, and Rappe]{Tan17}
Tan,~L.~Z.; Zheng,~F.; Rappe,~A.~M. {Intermolecular Interactions in Hybrid
  Perovskites Understood from a Combined Density Functional Theory and
  Effective Hamiltonian Approach}. \emph{ACS Energy Lett.} \textbf{2017},
  \emph{2}, 937--942\relax
\mciteBstWouldAddEndPuncttrue
\mciteSetBstMidEndSepPunct{\mcitedefaultmidpunct}
{\mcitedefaultendpunct}{\mcitedefaultseppunct}\relax
\EndOfBibitem
\bibitem[Lee \latin{et~al.}(2016)Lee, Lee, Kong, and Jang]{Lee16}
Lee,~J.~H.; Lee,~J.-H.; Kong,~E.-H.; Jang,~H.~M. {The nature of
  hydrogen-bonding interaction in the prototypic hybrid halide perovskite,
  tetragonal CH$_3$NH$_3$PbI$_3$}. \emph{Sci. Rep.} \textbf{2016}, \emph{6},
  21687\relax
\mciteBstWouldAddEndPuncttrue
\mciteSetBstMidEndSepPunct{\mcitedefaultmidpunct}
{\mcitedefaultendpunct}{\mcitedefaultseppunct}\relax
\EndOfBibitem
\bibitem[Poglitsch and Weber(1987)Poglitsch, and Weber]{Poglitsch87}
Poglitsch,~A.; Weber,~D. Dynamic disorder in methylammoniumtrihalogenoplumbates
  (II) observed by millimeter‐wave spectroscopy. \emph{J. Chem. Phys.}
  \textbf{1987}, \emph{87}, 6373--6378\relax
\mciteBstWouldAddEndPuncttrue
\mciteSetBstMidEndSepPunct{\mcitedefaultmidpunct}
{\mcitedefaultendpunct}{\mcitedefaultseppunct}\relax
\EndOfBibitem
\bibitem[Weller \latin{et~al.}(2015)Weller, Weber, Henry, Pumpo, and
  Hansen]{Weller15}
Weller,~M.~T.; Weber,~O.~J.; Henry,~P.~F.; Pumpo,~A. M.~D.; Hansen,~T.~C.
  Complete structure and cation orientation in the perovskite photovoltaic
  methylammonium lead iodide between 100 and 352 K. \emph{Chem. Commun.}
  \textbf{2015}, \emph{51}, 4180--4183\relax
\mciteBstWouldAddEndPuncttrue
\mciteSetBstMidEndSepPunct{\mcitedefaultmidpunct}
{\mcitedefaultendpunct}{\mcitedefaultseppunct}\relax
\EndOfBibitem
\bibitem[Leguy \latin{et~al.}(2016)Leguy, Go{\~n}i, Frost, Skelton, Brivio,
  Rodr{\'i}guez-Mart{\'i}nez, Weber, Pallipurath, Alonso, Campoy-Quiles,
  Weller, Nelson, Walsh, and Barnes]{Leguy16}
Leguy,~A. M.~A.; Go{\~n}i,~A.~R.; Frost,~J.~M.; Skelton,~J.; Brivio,~F.;
  Rodr{\'i}guez-Mart{\'i}nez,~X.; Weber,~O.~J.; Pallipurath,~A.; Alonso,~M.~I.;
  Campoy-Quiles,~M.; Weller,~M.~T.; Nelson,~J.; Walsh,~A.; Barnes,~P. R.~F.
  Dynamic disorder, phonon lifetimes, and the assignment of modes to the
  vibrational spectra of methylammonium lead halide perovskites. \emph{Phys.
  Chem. Chem. Phys.} \textbf{2016}, \emph{18}, 27051--27066\relax
\mciteBstWouldAddEndPuncttrue
\mciteSetBstMidEndSepPunct{\mcitedefaultmidpunct}
{\mcitedefaultendpunct}{\mcitedefaultseppunct}\relax
\EndOfBibitem
\bibitem[Marchenko \latin{et~al.}(in press)Marchenko, Fateev, Petrov, Korolev,
  Mitrofanov, Petrov, Goodilin, and Tarasov]{Marchenko20}
Marchenko,~E.~I.; Fateev,~S.~A.; Petrov,~A.~A.; Korolev,~V.~V.; Mitrofanov,~A.;
  Petrov,~A.~V.; Goodilin,~E.~A.; Tarasov,~A.~B. Database of 2D hybrid
  perovskite materials: open-access collection of crystal structures, band gaps
  and atomic partial charges predicted by machine learning. \emph{Chem. Mater.}
  \textbf{in press}, \relax
\mciteBstWouldAddEndPunctfalse
\mciteSetBstMidEndSepPunct{\mcitedefaultmidpunct}
{}{\mcitedefaultseppunct}\relax
\EndOfBibitem
\bibitem[Zhai \latin{et~al.}(2017)Zhai, Baniya, Zhang, Li, Haney, Sheng,
  Ehrenfreund, and Vardeny]{Zhai17}
Zhai,~Y.; Baniya,~S.; Zhang,~C.; Li,~J.; Haney,~P.; Sheng,~C.-X.;
  Ehrenfreund,~E.; Vardeny,~Z.~V. Giant Rashba splitting in 2D
  organic-inorganic halide perovskites measured by transient spectroscopies.
  \emph{Sci. Adv.} \textbf{2017}, \emph{3}, e1700704\relax
\mciteBstWouldAddEndPuncttrue
\mciteSetBstMidEndSepPunct{\mcitedefaultmidpunct}
{\mcitedefaultendpunct}{\mcitedefaultseppunct}\relax
\EndOfBibitem
\bibitem[Kilina \latin{et~al.}(2009)Kilina, Ivanov, and Tretiak]{Kilina09}
Kilina,~S.; Ivanov,~S.; Tretiak,~S. Effect of Surface Ligands on Optical and
  Electronic Spectra of Semiconductor Nanoclusters. \emph{J. Am. Chem. Soc.}
  \textbf{2009}, \emph{131}, 7717--7726\relax
\mciteBstWouldAddEndPuncttrue
\mciteSetBstMidEndSepPunct{\mcitedefaultmidpunct}
{\mcitedefaultendpunct}{\mcitedefaultseppunct}\relax
\EndOfBibitem
\bibitem[Kresse and Furthm{\"u}ller(1996)Kresse, and Furthm{\"u}ller]{Kresse96}
Kresse,~G.; Furthm{\"u}ller,~F. {Efficient iterative schemes for \textit{ab
  initio} total-energy calculations using a plane-wave basis set}. \emph{Phys.
  Rev. B} \textbf{1996}, \emph{54}, 11169--11186\relax
\mciteBstWouldAddEndPuncttrue
\mciteSetBstMidEndSepPunct{\mcitedefaultmidpunct}
{\mcitedefaultendpunct}{\mcitedefaultseppunct}\relax
\EndOfBibitem
\bibitem[Kresse and Joubert(1999)Kresse, and Joubert]{Kresse99}
Kresse,~G.; Joubert,~D. {From ultrasoft pseudopotentials to the projector
  augmented-wave method}. \emph{Phys. Rev. B} \textbf{1999}, \emph{59},
  1758--1775\relax
\mciteBstWouldAddEndPuncttrue
\mciteSetBstMidEndSepPunct{\mcitedefaultmidpunct}
{\mcitedefaultendpunct}{\mcitedefaultseppunct}\relax
\EndOfBibitem
\bibitem[Monkhorst and Pack(1976)Monkhorst, and Pack]{Monkhorst76}
Monkhorst,~H.~J.; Pack,~J.~D. {Special points for Brillouin-zone integrations}.
  \emph{Phys. Rev. B} \textbf{1976}, \emph{13}, 5188--5192\relax
\mciteBstWouldAddEndPuncttrue
\mciteSetBstMidEndSepPunct{\mcitedefaultmidpunct}
{\mcitedefaultendpunct}{\mcitedefaultseppunct}\relax
\EndOfBibitem
\bibitem[Klime{\v s} \latin{et~al.}(2010)Klime{\v s}, Bowler, and
  Michaelides]{Klimes10}
Klime{\v s},~J.; Bowler,~D.~R.; Michaelides,~A. {Chemical accuracy for the van
  der Waals density functional}. \emph{J. Phys.: Condens. Matter}
  \textbf{2010}, \emph{22}, 022201\relax
\mciteBstWouldAddEndPuncttrue
\mciteSetBstMidEndSepPunct{\mcitedefaultmidpunct}
{\mcitedefaultendpunct}{\mcitedefaultseppunct}\relax
\EndOfBibitem
\bibitem[Klime{\v s} \latin{et~al.}(2011)Klime{\v s}, Bowler, and
  Michaelides]{Klimes11}
Klime{\v s},~J.; Bowler,~D.~R.; Michaelides,~A. {Van der Waals density
  functionals applied to solids}. \emph{Phys. Rev. B} \textbf{2011}, \emph{83},
  195131\relax
\mciteBstWouldAddEndPuncttrue
\mciteSetBstMidEndSepPunct{\mcitedefaultmidpunct}
{\mcitedefaultendpunct}{\mcitedefaultseppunct}\relax
\EndOfBibitem
\bibitem[Men{\'e}ndez-Proupin \latin{et~al.}(2014)Men{\'e}ndez-Proupin,
  Palacios, Wahn{\'o}n, and Conesa]{Menendez-Proupin14}
Men{\'e}ndez-Proupin,~E.; Palacios,~P.; Wahn{\'o}n,~P.; Conesa,~J.~C.
  {Self-consistent relativistic band structure of the CH$_3$NH$_3$PbI$_3$
  perovskite}. \emph{Phys. Rev. B} \textbf{2014}, \emph{90}, 045207\relax
\mciteBstWouldAddEndPuncttrue
\mciteSetBstMidEndSepPunct{\mcitedefaultmidpunct}
{\mcitedefaultendpunct}{\mcitedefaultseppunct}\relax
\EndOfBibitem
\bibitem[Perdew \latin{et~al.}(1996)Perdew, Burke, and Ernzerhof]{Perdew96}
Perdew,~J.~P.; Burke,~K.; Ernzerhof,~M. {Generalized Gradient Approximation
  Made Simple}. \emph{Phys. Rev. Lett.} \textbf{1996}, \emph{77},
  3865--3868\relax
\mciteBstWouldAddEndPuncttrue
\mciteSetBstMidEndSepPunct{\mcitedefaultmidpunct}
{\mcitedefaultendpunct}{\mcitedefaultseppunct}\relax
\EndOfBibitem
\end{mcitethebibliography}


\end{document}